\documentclass[oneside,onecolumn,12pt,a4paper]{article}

\usepackage{graphicx}
\usepackage{dsfont}
\usepackage{natbib}
\usepackage{bbding}
\usepackage{setspace}
\usepackage{braket}
\usepackage{sectsty}
\usepackage[multiple]{footmisc}
\usepackage[margin=30mm]{geometry}
\usepackage{tikz}
\usepackage{bm}
\usepackage{authblk}
\usepackage{amsmath}
\usepackage{lmodern}
\usepackage{commath}
\usepackage{mathtools}
\usepackage{amssymb}
\usepackage{hyperref}
\usepackage{qtree}
\usepackage{combelow}
\usepackage{csquotes}
\usepackage{color}
\definecolor{darkblue}{rgb}{0.0,0.0,0.3}
\hypersetup{colorlinks,breaklinks,linkcolor=darkblue,urlcolor=darkblue,anchorcolor=darkblue,citecolor=darkblue}

\usetikzlibrary{calc}
\usetikzlibrary{arrows}
\usetikzlibrary{decorations.markings}
\usetikzlibrary{shapes}
\usetikzlibrary{positioning}

\usepackage[justification=centering]{caption}
\usepackage[labelformat=simple]{subcaption}

\usepackage{ifthen}
\def\eprinttmp@#1arXiv:#2 [#3]#4@{\ifthenelse{\equal{#3}{}}{\href{http://arxiv.org/abs/#1}{arXiv:#1}}{\href{http://arxiv.org/abs/#2}{arXiv:#2 [#3]}}}
\newcommand{\eprint}[1]{\eprinttmp@#1arXiv: []@}
\newcommand{\doi}[1]{\href{http://dx.doi.org/#1}{doi:#1}}

\sectionfont{\normalfont\normalsize\bfseries}
\subsectionfont{\normalfont\normalsize\bfseries}
\subsubsectionfont{\normalfont\normalsize\bfseries}

\begin{document}
\sloppy
\setstretch{1.2}

\title{On the Limits of Experimental Knowledge\footnote{Forthcoming in \textit{Philosophical Transactions of the Royal Society A}, \textit{The Next Generation of Analogue Gravity Experiments}, edited by Maxime Jacquet, Silke Weinfurtner and Friedrich K\"onig.}}

\author[1]{\bf Peter W. Evans\thanks{email: \href{mailto:p.evans@uq.edu.au}{p.evans@uq.edu.au}}}
\author[2]{\bf Karim P. Y. Th\'ebault\thanks{email: \href{mailto:karim.thebault@bristol.ac.uk}{karim.thebault@bristol.ac.uk}}}

\affil[1]{\small{{\it School of Historical and Philosophical Inquiry}, University of Queensland}}
\affil[2]{\small{{\it Department of Philosophy}, University of Bristol}}
	\maketitle

\begin{abstract}
	To demarcate the limits of experimental knowledge we probe the limits of what might be called an experiment. By appeal to examples of scientific practice from astrophysics and analogue gravity, we demonstrate that the reliability of knowledge regarding certain phenomena gained from an experiment is not circumscribed by the manipulability or accessibility of the target phenomena. Rather, the limits of experimental knowledge are set by the extent to which strategies for what we call `inductive triangulation' are available: that is, the validation of the mode of inductive reasoning involved in the source-target inference via appeal to one or more distinct and independent modes of inductive reasoning. When such strategies are able to partially mitigate reasonable doubt, we can take a theory regarding the phenomena to be well supported by experiment. When such strategies are able to fully mitigate reasonable doubt, we can take a theory regarding the phenomena to be established by experiment. There are good reasons to expect the next generation of analogue experiments to provide genuine knowledge of unmanipulable and inaccessible phenomena such that the relevant theories can be understood as well supported.
\end{abstract}

\newpage

\tableofcontents

\section{Introduction} 

It is somewhat of an oversimplification to say that experiments allow us to gain knowledge about the world. Indeed, an experiment, in and of itself, may not allow us to gain any reliable knowledge at all. Consider a measurement of negative temperature with a faulty digital thermometer or the infamous detection of neutrinos moving at superluminal speed at OPERA. In and of itself, an experiment need not teach us anything useful, even about the system that is being directly manipulated. For us to gain reliable knowledge from an experiment, it must be the case that the experiment is validated. Two distinct forms of experimental validation are differentiated by the object system about which we are justified in believing we have gained knowledge. An \textit{internally} valid experiment justifies our beliefs about a \textit{source} system, which is directly manipulated in the experiment. An \textit{externally} valid experiment justifies our beliefs about a \textit{target} system, which is not directly manipulated in the experiment.\footnote{This concept of `external validity' is closely analogous to that used in the social sciences.} Typically the internal validity of a given experiment is necessary but not sufficient for the external validity of that experiment.

What kinds of target systems can we gain knowledge about? And what factors place limits on the strength of this knowledge? In particular, must target systems be, in principle, themselves \textit{manipulable}? Or should we insist that they are at least \textit{accessible}, in the sense of being subject to possible observation? In this paper we will argue that the limits of experimental knowledge should not be taken to be circumscribed by the manipulability or accessibility of target systems. There is no, in principle, epistemic barrier to experiments with unmanipulable or inaccessible target systems being externally valid. Experiments in contemporary science can and do allow us to gain knowledge of unmanipulable and inaccessible target systems. We will argue that the limits of experimental knowledge are in fact set by the \textit{mitigation of reasonable doubt} -- that is, the application of inductive strategies for internally and externally valid source-target inferences. When reasonable doubt has been partially mitigated, a theory can be said to be \textit{well supported}, and a scientist is justified to treat the empirical consequences of the theory as likely to be true, in the relevant domain. When reasonable doubt has been almost entirely mitigated a theory can be said to be \textit{established}, and a scientist is justified to treat the empirical consequences of the theory as true, in the relevant domain. To demarcate the limits of experimental knowledge we will probe the limits of what might be called an experiment. In particular, we will illustrate our arguments by drawing upon examples from astrophysics and analogue gravity, in the latter case drawing upon existing philosophical work on the epistemology of analogue black holes \citep{Dardashti:2015,Dardashti:2019,Thebault:2016}. Using these examples we will illustrate three core claims.

Our first core claim is that whether a theory regarding certain phenomena can be well supported or established by experiment is not constrained by the requirement that the target system displaying these phenomena be manipulable or accessible, either in principle or practice. Thus, theories regarding unmanipulable and inaccessible phenomena can in principle become established via experiment. We thus endorse a liberal form of empiricism within which the scope of phenomena about which we can gain experimental knowledge is much wider than that of either restricted forms of empiricism \citep{vanFraassen:1980,sep-constructive-empiricism} or, moreover, `detectionist' forms of scientific realism \citep{azzouni:2004,chakravartty:2007}. On our view, the limits of experimental knowledge are set by the extent to which strategies for \textit{inductive triangulation} are available: that is, the validation of the mode of inductive reasoning involved in the source-target inference via appeal to one or more distinct and independent modes of inductive reasoning. Our view is thus in direct opposition to the seemingly commonsensical view that the limits of experimental knowledge \text{are} set by what we can observe and manipulate. In order to convince the reader of our conclusion, we will draw upon a detailed study of inferences involved in contemporary scientific practice. In particular, we will show that nuclear process in the stellar core provide a vivid example of an unmanipulable and (at least partially) inaccessible target phenomenon the modern theories of which we can take to be uncontroversially established by conventional experiment.

Our second core claim is that an experiment on a manipulable and accessible source system of a given type can, in some circumstances, be used to make inductive inferences regarding a target system of a different type. In particular, when we have good reason to believe that the source and target phenomena in question are `universal', it is possible to make inductive inferences from the existence of the phenomena in the source system to the existence of the phenomena in the target system. Such inferences constitute a new form of `inter-type' uniformity principle which play a structurally similar role within the relevant inductive inference to that played by more standard uniformity principles (space, time, intra-type) in conventional inductive reasoning. As such, they are amenable to inductive triangulation. That is, we can justify source-target inferences based upon inter-type uniformity by appeal to one or more distinct and independent modes of inductive reasoning. Thus there is nothing in principle that rules out inter-type inductive reasoning from playing a significant role in inductive triangulation.

One third and final core claim is that analogue gravity experiments instantiate an example of scientific practice in which a scientist can use inductive triangulation to gain \textit{genuine knowledge} of inaccessible target phenomena based upon a \textit{highly speculative but structurally sound} inductive inference between accessible and inaccessible systems of different types. In particular, we will show how inductive triangulation can be applied such that a series of analogue black hole experiments that exploit multiple different types of source system can be combined with universality arguments to provide inductive evidence regarding the phenomena of Hawking radiation in astrophysical black holes. The explicit appeal to inductive triangulation provides concrete refutation of the claim appearing in the philosophical literature that `analogue experiments are not capable of confirming the existence of particular phenomena in inaccessible target systems' \cite[p.~1]{crowther:2019} due to a supposed `circularity' within the relevant chain of reasoning. These `circularity' arguments will be considered in detail in the final section and it will be shown that the only plausible reading of `circularity' in the given context is `rule circularity'. This form of circularity arises when one employs an argument to establish a proposition concerning a rule of inference, and the relevant argument-form employed is an instance of that rule. However, so long as one is not arguing with an inductive sceptic, inductive triangulation provides precisely the means to deflate such putative problems with rule-circularity, both in general and in the case of the highly speculative inferences from analogue gravity experiments to Hawking radiation in black holes. Inductive triangulation can thus, in principle, be extended as a means to cross-validate inferences that rely upon \textit{both} uniformity between accessible and inaccessible systems \textit{and} systems of different types.

There is thus a plausible epistemic basis to believe that the next generation of analogue experiments may be able to provide genuine knowledge of unmanipulable and inaccessible phenomena such that the relevant theories can be understood as well supported. Furthermore, looking further to the future, inductive triangulation allows for the possibility of analogue experiments to play a role, when combined with appropriate conventional experimental results, in establishing new theories.

\section{Epistemology and Experiment} 
\subsection{Reasonable and Unreasonable Doubt}

An ampliative inference is one in which the conclusion goes beyond what is (logically) entailed by the premises: it is not logically necessary that the conclusion is true given the truth of the premises. Inductive inferences can be defined as the set of all inferences that are ampliative.\footnote{Here we are taking `ampliative' and `inductive' to thus be synonymous. We mention both terms here, despite their synonymy, since some authors utilise `inductive' in a more narrow sense, as referring to an ampliative inference of a particular form.} Most scientists and philosophers -- with the notable exception of \cite{popper:1959} -- would hold that empirical science is based upon inductive inference (usually in combination with deductive inference). The Scottish Enlightenment philosopher David Hume \citep{hume:2016} famously identified the problem of finding a non-circular justification for inductive reasoning, known as `the problem of induction'. Hume argued that inductive reasoning must always assume that instances of which we have had no experience must resemble those of which we have had experience. This in turn, according to Hume, relies upon \textit{the} principle of the uniformity of nature, according to which there is similarity or resemblance between observed and unobserved regularities in nature \citep{Henderson:2018}. Hume's crucial observation was that in justifying such a principle we inevitably require further inductive reasoning. We are thus required to engage in a circular form of reasoning in justifying induction via induction itself.

Within the vast literature on the problem of induction \citep{Carnap:1952,Braithwaite:1953,goodman1954,Salmon_1963,CLEVE_1984,sober:1988,Papineau_1992,Okasha_2001,Norton_2003,Okasha_2005,psillos:2005,bird:2010,sep-abduction,Henderson:2018,schurz:2019} two lines of response will be of particular relevance for our discussion. First, it has been pointed out \citep{sober:1988,Norton_2003,Okasha_2001,Okasha_2005} that the assumption of a single principle of the uniformity underlying all inductive inference is at odds with actual scientific practice. To formulate a descriptively adequate argument, an inductive sceptic such as Hume must recognise the diversity of uniformity principles at work in our inductive practices: for example, temporal uniformity (past phenomena resemble future phenomena), spatial uniformity (local phenomena resemble distant phenomena), intra-type uniformity (these electrons resemble other electrons), and a huge variety of mixed uniformity principles that are based upon combinations of the others (stars inside our causal past resemble stars outside our causal past). It can then be argued that this recognition of diversity in uniformity principles serves to convert the circularity problem into a justificatory regress problem; since the various distinct uniformity principles must themselves be justified. Now, whether or not converting the problem from one of circularity to one of regress can be seen as a victory against the sceptic is not at all clear \citep{schurz:2019}. However, what is undoubtedly the case is that this line of response highlights the descriptive necessity of recognising the diversity of uniformity principles that underlie inductive inferences in actual science.

The second line of response \citep{CLEVE_1984,Papineau_1992} centres on the idea that we can distinguish between two notions of circularity to dissolve Hume's problem: `premise circularity' and `rule circularity' \citep{Carnap:1952,Braithwaite:1953,psillos:2005,bird:2010,sep-abduction}. Premise circularity (or begging the question) occurs when the conclusion of an argument is explicitly listed amongst the premises. Premise circular arguments are always \textit{viciously circular}, in the sense that putting forward a premise circular argument always involves making an informal fallacy of reasoning. Moreover, premise circular arguments are always dialectically ineffective in that they cannot be deployed to rationally convince an opponent of the truth of their conclusion. If Hume had shown that any inductive justification of induction were premise circular, then there would be a serious problem. However, so the counter-argument goes, in fact the justification of induction is properly thought of as `rule circular' rather than premise circular, and rule circularity is not always vicious. Rule circularity arises when one employs an argument to establish a proposition concerning a rule, such as its reliability, and the relevant argument-form towards the proposition is an instance of that same rule. An argument for the reliability of a given rule that essentially relies on the rule as an inferential principle is not viciously circular, provided that the use of the rule does not guarantee a positive conclusion about the rule's reliability. That is, rule circular arguments towards the reliability of a given rule do not constitute informal fallacies of reasoning (analogous to begging the question) unless they make their own reliability a sure thing. We can therefore see that rule circular inductive inferences cannot by definition be viciously rule circular since, as ampliative inferences, they cannot guarantee a positive conclusion about their own reliability. The crucial question is then whether the inductive justification of induction is dialectically ineffective or not. That is, granted that it cannot be viciously rule circular, there is still the question of whether an inductive justification of induction can be deployed to rationally convince an opponent.

What is crucial here is the dialectical context. If a particular instance of successful inductive inference is used to justify a general mode of inductive inference of the same form against a sceptical argument like Hume's that is based upon a single uniformity principle, then the rule circularity undermines the dialectic force of such an argument: it gives no reason for an inductive sceptic to change their mind regarding the point at dispute. However, if a particular instance of inductive inference is employed within an argument to justify a second (non-identical) instance of inductive inference, then the argument may well have dialectic force against an interlocutor who is not sceptical of inductive reasoning \emph{per se}.\footnote{Our analysis of rule circularity and inductive reasoning is very similar in spirit to that given by \cite{carter:2017} in the context of a discussion of rule circularity and inference to the best explanation. In particular, the potentially dialectically convincing example of a rule circular inductive argument we are considering here is an instance of what they call a `wide' rather than `narrow' rule circular argument. On their analysis the former, but not the latter, have justificatory structure that is not necessarily defective.} Consider the example of using inductive arguments based upon the temporal uniformity of nature to justify reasoning based upon spatial uniformity: in the past, distant observed phenomena have regularly resembled local observed phenomena, so in the future distant unobserved phenomena will resemble local observed phenomena. Here one inductive inference (based on, say, particular observations of spatial regularities) is reinforced by an independently established inductive inference (in this case, temporal uniformity). Or consider using inductive arguments based upon spatial uniformity to justify an inductive argument for uniformity between different tokens of the same type: the properties of observed spatially distant electrons resemble the properties of observed local electrons, so all unobserved electrons will resemble observed electrons. Here, again, one inductive inference (based on, say, particular observations of regular properties across tokens of some type) is reinforced by an independently established inductive inference (in this case, spatial uniformity). Clearly such arguments can be, and often are, convincing.

Let us call such a style of reasoning \textit{inductive triangulation}. As the name suggests, this idea has much in common with the idea of `triangulation' \citep{feigl:1958} that has been discussed in the context of the social and, particularly, the historical sciences \citep{webb:1966,wylie:2002,chapman:2014,currie:2018}. Our usage is, however, somewhat more specific since it relates to distinct modes of inductive reasoning, rather than simply distinct lines of evidence. Plausibly, if inductive triangulation is deployed with the specific aim of defeating the inductive sceptic then it allows one to avoid any dialectically problematic rule circularity. However, once more, an obvious problem of infinite regress looms large: each link in the chain of inductive inferences can be justified in a non-circular way, but it is not clear how to terminate the regress of justificatory demands. Again, it is unclear whether or not a response based upon rule circularity and inductive triangulation improves the situation \emph{vis-a-vis} the problem of inductive scepticism \citep{skyrms:2000,schurz:2019}. However, what is clear is that there are convincing means by which to justify novel forms of inductive reasoning against a sceptic who is not globally sceptical of induction. That is, if inductive triangulation is deployed with the aim of convincing someone to extend the licensed forms of inductive inference then there is no dialectally problematic rule circularity.

The lesson is that, provided both parties to a dispute regarding the reliability of a particular mode of inductive inference accept some form of inductive reasoning then inductive triangulation is a legitimate (yet defeasible) means to establish justification. An inductive justification of induction is not always dialectically problematic. In a scientific context it is simply unreasonable not to admit any form of inductive reasoning, and thus inductive triangulation is always an admissible argumentative strategy. This leads us to define \textit{unreasonable doubt}, in a scientific context, as doubt regarding the reliability of a specific instance of inductive reasoning that cannot be mitigated via further inductive reasoning, including inductive triangulation. We can then define \textit{reasonable doubt}, in a scientific context, as doubt regarding the reliability of a specific instance of inductive reasoning that can be mitigated via further inductive reasoning, including inductive triangulation.

\subsection{Three Forms of Unobservable Phenomena}

Once the spectre of unreasonable doubt has been clearly distinguished from its reasonable counterpart, a constructive philosophical analysis of inductive practices in science can be pursued in isolation from Hume's problem. The question of particular relevance is the relationship between the observed and the unobserved. In particular, can we find strategies for inductive triangulation to validate such inferences. To pose this question precisely we will require a number of further distinctions.

The first and most basic is between the data gleaned from a particular experiment or observation and the general class of observable phenomena about which scientists may reasonably draw conclusions, given the data. Consider the canonical exemplar of Galileo's observation of the phases of Venus: the data would be the particular spots of light that Galileo saw through his telescope and the observable phenomena would be the phases of Venus themselves. In principle both of these are observable in the sense of visually accessible. In general, there being no need to privilege sight above the other senses, we can think of observables as physical quantities whose value can be directly discerned via the senses. The important difference here between data and phenomena is that the data are idiosyncratic to a specific experimental context but the phenomena are not \citep{bogen:1988}. In the case we are considering here, both the data and phenomena are observable but, while the data is actually observed, the observable phenomena are not. We can have reasonable inductive doubts about both data and phenomena. What if Galileo's telescope was faulty? What if he had observed Venus in an atypical part of its orbit? In each case such reasonable doubts are mitigated precisely by inductive triangulation: testing the telescope on different celestial objects, re-observing the phases of Venus at a different time of the year.

\begin{figure}
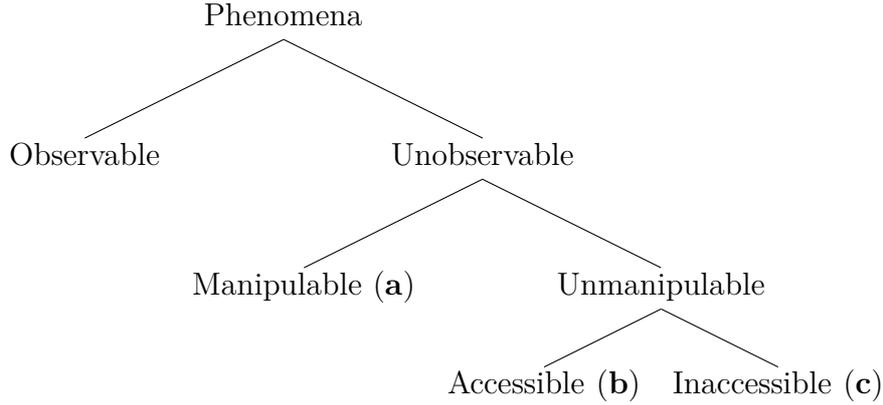

\Tree [.Phenomena Observable [.Unobservable Manipulable\space(\textbf{a}) [.Unmanipulable Accessible\space(\textbf{b}) Inaccessible\space(\textbf{c}) ] ] ] 
\caption{Diagram Illustrating the three forms of unobservable phenomena.}
\label{fig:3p}
\end{figure}

Whilst observable phenomena were often indeed the focus of Renaissance astronomy, most of modern science is built upon inferences regarding unobservable phenomena. In particular, as powerfully argued by \cite{massimi:2007}, building on the original work on data and phenomena due to \cite{bogen:1988}, such unobservable phenomena are the subject of almost all experimental practice in modern particle physics. The main focus of this section is to differentiate three different types of unobservable phenomena (see Fig. \ref{fig:3p}). The first, and most basic, are unobservable phenomena that are manipulable. Consider another canonical example: the Stern-Gerlach experiment. Here the data are spots on a particular screen and the phenomenon is the spin of the electron -- that is, the measurable quantity corresponding to spin rather than the theoretical concept spin. This is an unobservable phenomenon in the sense that it is not a physical quantity whose value can be directly discerned via the senses. However, the spin of the electron clearly is a physical quantity whose value can be indirectly discerned. Moreover, although we cannot of course change the numerical value of the electron spin, it is a vector quantity and via experimental apparatus like the Stern-Gerlach set-up we can change the orientation of the spin. The manipulation of (tokens of) the relevant unobservable phenomena is in turn an important part of the story about how, again via inductive triangulation, we can mitigate reasonable doubts regarding the inferences from experimental data to unobservable phenomena.

In general terms, unobservable phenomena that are manipulable (\textbf{a}) correspond to phenomena to which we have `two way' causal access. That is, we can probe the phenomena via a suitable mediating system, and the phenomena can `push-back' via such a system. Whilst much of modern physical science does indeed focus on such phenomena, it would be premature to terminate our analysis here. Rather, moving beyond particle physics into the realm of astrophysics and cosmology it is obviously the case that the unobservable phenomena of interest are  unmanipulable. Usually this is because they are very far away, happened a long time ago, or are simply far too big. The phenomenon of a black hole merger as detected via gravitational waves is perhaps the most vivid recent example of unobservable, unmanipulable phenomena but it is not difficult to come up with a host of other examples.\footnote{Three further examples are: the internal structure and composition of the Earth's core as determined by measurements at the Earth's surface of type P and S seismic waves; the existence of exoplanets as determined by measurements of radial velocity of stars with respect to the Earth; and the value of the cosmological constant. In each case there is an unobservable, unmanipulable phenomenon that can be experimentally quantified to a large degree independently of the relevant theoretical interpretation regarding their nature. For an extensive discussion of the question of the `authentication' of empirical phenomena, and theoretical perspectives on the nature of such phenomena in the context of the history of particle physics, see \citep{cretu:2020}.} The story about how our inductive inferences about such phenomena are validated is often a more complex one than in the case of manipulable phenomena. However, it is noteworthy that, once more, scientists can and do employ a wider variety of inductive triangulation strategies.

Unobservable, unmanipulable phenomena can themselves be further differentiated on the basis of whether or not we have `one way' causal access or not. That is, whether or not such phenomena have discernible physical effects on observable systems to which we have access. We thus have two further forms of unobservable phenomena, those that are unmanipulable and accessible  (\textbf{b})  and those that are unmanipulable and inaccessible (\textbf{c}). Black hole mergers are accessible in the relevant sense. Examples of  phenomena that are inaccessible in principle, at least according to current physics, include the physics of black holes behind the event horizon and all physical phenomena outside our cosmological particle horizon \citep{davis:2004}. There are also, of course, examples of phenomena that are physically inaccessible in practice. The two examples that will be discussed in detail in this paper are the photonic physics of stellar nucleosynthesis and Hawking radiation associated with black hole event horizons. In such cases the relevant phenomena are in principle accessible, however the relevant signal is vanishingly small and so, in practice, \textit{according to current physics}, we are never likely to be able to access them. There is thus a strong time-dependence to the notion of `in practice' inaccessibility which introduces a degree of vagueness. Here we will use a working definition of in practice inaccessible phenomena as phenomena that we have good physical reasons to believe will remain inaccessible within the framework of contemporary science and near future technology.

How can we ever expect to learn about in practice inaccessible phenomena through observation or experiment? Would not inductive inferences regarding such phenomena always be subject to a quite devastating and reasonable form of doubt? How can we construct inductive triangulation procedures to mitigate such doubt? Such questions will be taken up in Sections \ref{sec:stellar} and \ref{sec:Hawking} in the context of the examples of stellar nucleosynthesis and Hawking radiation. Before then we must provide a final piece of philosophical machinery: an analysis of confirmation and evidence in the context of contemporary experimental science.

\subsection{Experimental Evidence and External Validation}

Our focus here is on the factors that influence the strength of support that experimental evidence can offer a theory or model describing unobservable phenomena. In all cases, what is crucial to the strength of the relevant inferences is an extrapolation from a manipulated system that is the subject of the experiment (`the source') to a further class of unmanipulated (but in some cases potentially manipulable) systems that display the relevant unobservable phenomena (`the target'). A simple example is given by experiments designed to learn about the iron content in the core of the Earth by superheating a sample of iron in a lab using lasers \citep{konopkova:2016,dobson:2016}. The experiments are carried out in the lab using samples of iron that are placed in a laser-heated diamond-anvil cell. The pressure and temperature to which the iron samples are subjected are specifically matched to those relevant to the cores of Mercury-sized to Earth-sized planets. Iron in the core of Mercury-sized to Earth-sized planets is the target, the iron in the lab is the source. Consider a particular theory of geophysics set out in terms of the predicted phenomenon of the thermal conductivity displayed by the iron in the core. In what circumstances can we take evidence regarding thermal conductivity drawn from the experiment on the source system to support a theory regarding the target phenomenon? And what determines the strength of the support? Such questions are usually posed in terms of the idea of external validation, which will be worthwhile discussing in some detail.

After a long period of relative neglect, the philosophy of experimental physics is now the subject of an extensive literature \citep{hacking:1983,galison:1987,franklin:1989,sep-physics-experiment}. One of the most significant points established in such discussions is that, in and of itself, an experiment need not teach us anything, even about the system that is being directly manipulated. Rather, an experiment is only genuinely probative of the system that is being experimented on when it has been \textit{internally validated} through the establishment of, for instance, the reliability of the apparatus and the robustness of the experimental protocol. Such a process of internal validation can be understood precisely in terms of the idea of mitigation of reasonable doubt discussed earlier. In practice many elements of the internal validation of an experiment take the form of explicit `auxiliary hypotheses'; statements relating to instrumentation or other background assumptions that are needed to support any inferences from the observational statements gained through the experiment. Also significant is the role of more practical, sometimes tacit, experimental knowledge in internal validation. Experimenters are embedded in a scientific tradition that includes complex protocols for conducting a given experiment type in a valid manner. Finally, in addition to auxiliary hypotheses and experimental tradition, internal validation typically involves some consideration of statistical error. That is, when the system being experimented on is assumed to be stochastic or subject to random external fluctuations, experimenters need to establish that the properties attributed to it are, to the relevant degree, typical of its stable state, rather than statistical aberrations. For the most part this form of validation is very difficult to achieve for a single system. Thus, multiple suitably similar systems are experimented upon. In each case the strategies for achieving internal validation, and thus mitigating reasonable doubt, are inductive. However, they need not involve inductive triangulation. That is, the same form of inductive reasoning that is involved in making inferences about the source system based upon the experimental data might be used to internally validate the experiment itself.

External validation is then mitigation of reasonable doubt regarding whether the source system is relevantly probative of the second, unmanipulated, target system or class of systems.  Similarly to internal validation, external validation involves a combination of auxiliary statements, often from well-established scientific theory, practical experimental knowledge, and statistical reasoning. However, unlike internal validation, in the case of external validation there is typically a requirement for inductive triangulation at the heart of the mitigation of the relevant reasonable doubt. The principal reason for this is that typically the source-target inference requires an appeal to intra-type uniformity: the experiment on the source system is taken to be relevantly probative of the target phenomena on the basis that they are tokens of the same type of substance. By what means can one mitigate reasonable doubt regarding the general pattern of such an inference? If one is confronted by a sceptic regarding inductive inferences based upon intra-type uniformity, how can one respond? In the context of such an opponent there is the obvious danger of rule circularity of a dialectically undermining sort.

Consider for instance justifying an inference from particular iron atoms in the source to iron atoms in the target based upon intra-type uniformity between the relevant nucleons and electrons. The pattern of inference which we are seeking to justify is now itself involved in the justificatory argument. The argument thus provides no dialectic force against the intra-type uniformity sceptic since its rule circularity means the sceptic has been given no extra reason to change their mind. The key point here is that, for the doubt to be reasonable in such circumstances, our interlocutor must admit to some forms of inductive reasoning. Thus we can mitigate general, reasonable doubt regarding intra-type uniformity in a dialectically convincing way by invoking inductive inferences built upon spatial or temporal uniformity. These electrons resemble other electrons \textit{because} there is an assumed spatial uniformity between local phenomena and distant phenomena. This sample of iron in the lab is like iron in the core of the Earth in the relevant respects \textit{because} past experiments and observations have been used to calibrate the relevant experimental parameters.

It is important to note here that the manipulability and accessibility of the target system does not in and of itself tell us anything about the limits to such external validation processes. There is nothing in principle that tells us that external validation for manipulable target systems is easier to achieve for accessible target systems nor, moreover, that such validation will even always be easier for accessible over inaccessible systems. The degree to which reasonable doubt can be mitigated via inductive triangulation depends upon contingent features specific to the experiment, the source and target phenomena in question, and various theoretical and historical circumstances.

We are now finally in a position to set out our stance regarding the limits of experimental knowledge. We take it that such limits are set by the mitigation of reasonable doubt -- that is, the availability of inductive strategies for internally and externally validating source-target inferences. When reasonable doubt has been partially mitigated a theory can be said to be \textit{well supported}, and a scientist is justified to treat the empirical consequences of the theory as likely to be true, in the relevant domain. Inductive triangulation may be required in such a mitigation process, but it also may not. When reasonable doubt has been almost entirely mitigated a theory can be said to be \textit{established}, and a scientist is justified to treat the empirical consequences of the theory as true, in the relevant domain. Plausibly, inductive triangulation will always be required in such a mitigation process. In neither case is the issue of inductive scepticism, and thus unreasonable doubt, relevant.\footnote{An outstanding problem for our analysis is the means by which the notions of a theory being `well supported' and `established' could be translated into probabilistic terms. In particular, it would be very plausible to attempt to connect the relation of confirmation, as formalised probabilistically within Bayesian confirmation theory, with the status of a theory as well supported or established. Plausibly, well-supported theories are those in which updating from reasonable priors based upon the evidence leads to a moderately high posterior probability (e.g. greater than 0.5) and established theories are those in which updating from reasonable priors based upon the evidence leads to a very high posterior probability (e.g. greater than 0.99). Unfortunately such a straightforward translation immediately leads to well-known problems with the definition of both reasonable priors (see e.g. \citep{salmon:1990}) and suitable thresholds for ``full belief'' (see e.g. \citep{buchak:2014}). Such complexities do not block the explanatory value of our analysis for scientific practice but would certainly need to be solved for the development of a fully fledged epistemology of experiment.}

Our notion of an established theory closely resembles what \cite{dawid:2019} calls  `conclusive confirmation', which he defines as when a ``theory has been established to be [empirically] viable in a given regime beyond reasonable doubt'' (p.~105). Whereas Dawid leaves reasonable doubt undefined, we propose to explicitly define reasonable doubt as doubt that is amenable to mitigation based upon inductive reasoning. Reasonable doubt in this sense obviously specifically excludes inductive doubt, but is also defined such that it excludes doubt based upon Cartesian scepticism or radical social constructivism regarding experimental  knowledge.\footnote{That at least some forms of social constructivism can be understood as unreasonable doubt is evidenced by, for instance, the sentiment that: ``a sufficiently determined critic can always find a reason to dispute any alleged \enquote{result}'' \citep{mackenzie:1989}. An instructive summary of the debates regarding constructivism about experimental knowledge is given in \S1.2 of \citep{sep-physics-experiment}. See also the excellent discussion of theory-laden experimentation due to  \cite{schindler:2013}.} Our point is not that such doubts are unreasonable \textit{per se}, but rather that they are unreasonable in the context of an analysis of the epistemology of actual scientific practice, a context in which the acceptance of at least some form of inductive reasoning is a methodological \textit{sine qua non}.\footnote{There is thus some similarity between what we are proposing and the `response' to Cartesian scepticism deployed by epistemic contextualism \citep{sep-contextualism-epistemology}.}

As already noted, the manipulability and accessibility of some target phenomena does not, on our view, in and of itself constrain the potential for the mitigation of reasonable doubt, and thus the potential for a theory regarding the phenomena to be well supported or established. In principle, it is thus perfectly possible for theories regarding inaccessible phenomena to be taken to be well supported or established based upon a suitably externally validated experiment and (where necessary) inductive triangulation strategy. We thus endorse a liberal form of empiricism within which the scope of phenomena about which we can gain experimental knowledge is \textit{in principle} much wider than that of either restricted forms of empiricism \citep{vanFraassen:1980,sep-constructive-empiricism} or, moreover, `detectionist' forms of scientific realism \citep{azzouni:2004,chakravartty:2007}. Whether and how such knowledge can be obtained \textit{in practice} will be the focus of the remainder of the paper.\footnote{For an excellent discussion highlighting the problems with detectionist forms of realism in the context of cosmological horizons and dark matter see \citep{Allzen:2020}. An example of a liberal empiricism with interesting similarities to our own is that according to Hintikka's remarkable exegesis of Mach's empiricism in terms of the econometricians notion of identifiability \citep{hintikka:2001}. An additional connection, which we hope to explore in future work, is between our notion of inductive triangulation and Massimi's idea that the deployment of a plurality of seemingly incompatible models is methodologically crucial to the establishment of knowledge claims in the context of contemporary particle physics \citep{massimi:2018}.}

\section{Case Study I: Stellar Nucleosynthesis} 
\label{sec:stellar}
\subsection{Evidence and Experiment}
Our first case study chosen to help demarcate the limits of experimental knowledge is the model of stellar nucleosynthesis -- that is, the model of the nuclear reactions that form the primary sources of energy production inside the core of a star. What makes this case study particularly salient in the context of our discussion is that it both involves all three forms of unobservable phenomena, (\textbf{a})--(\textbf{c}), and a process of external validation built upon inductive triangulation. Furthermore, the model of stellar nucleosynthesis is supported by validated experimental evidence of such quality and quantity that it is plausibly taken to be established. That is, relevant reasonable doubts have been almost entirely mitigated, and thus scientists are justified to treat the empirical consequences of the theory as true, in the relevant domain. Before we consider these epistemological claims in detail, let us consider the model of stellar nucleosynthesis as described in contemporary physics.

For main sequence stars the model of stellar nucleosynthesis consists of two principal reactions that take place in the stellar core: the proton-proton (pp) chain; and the carbon-nitrogen-oxygen (CNO) cycle \citep{Rose1998}. In stars such as our Sun, the pp chain is the dominant source of energy production, and transforms hydrogen, via deuterium, into helium, with energy released in the form of gamma rays. Less dominant in stars the size of our Sun (but becoming more dominant in larger stars), the CNO cycle also transforms hydrogen into helium, but does so via a catalytic process. In this process a carbon nucleus sequentially captures four protons. By this capture process, and two associated weak nuclear decays, the carbon nucleus is transformed to a nitrogen, and then an oxygen, nucleus before returning to carbon after releasing a helium nucleus, with the release of energy in the form of gamma rays at each step along the way.

The most significant feature of these processes for our purposes is that they occur deep within the stellar core. The high energy photons that result from these processes are released into the dense plasma of the stellar interior and so, due to their strong interaction with matter, have a mean free path of about the order of a centimetre. The origin of the stellar photons that we observe from the Earth then is always the stellar surface layers and thus, with regard to photons at least, processes going on within the interior of stars are in practice entirely inaccessible. Thus obtaining direct photonic observational evidence for the nuclear processes at the stellar core is simply not possible. As a result, these processes are unmanipulable, and (photonically) inaccessible. We thus have an example of target phenomena from the third most removed form of unobservable phenomena (\textbf{c}). Despite this, so we will argue below, the model describing such phenomena is so well supported by externally validated experimental evidence that there is little if any room for reasonable doubt. Scientists are thus justified in treating the empirical consequences of the theory as true in the relevant domain. Let us consider the various sources of experimental evidence in turn.

To begin with, any possible source of stellar energy production is constrained by two factors. Firstly, isotope abundances calculated from transition rates between isotopes in any putative process of energy production are constrained by the isotope abundances we observe in space, which themselves vary across `old' and `new' regions of the universe, and between stars and interstellar space. Secondly, the rate of reaction for any putative process of energy production is constrained by the inferred core temperatures and lifetimes of stars of different masses. We can get a better grasp on how these constraints restrict model possibilities by considering the role that they played in the development of the first light-element nuclear transitions proposed as the energy source of stars \citep{Gamow1935,Gamow1938}. These proposals transgressed against the constraints by either suggesting isotope abundances mismatched to observation -- in particular, interstellar abundances of lithium, beryllium, and helium isotopes -- or by containing reactions that are, based on known cross-sections, too rapid or too slow to match inferred stellar lifetimes. In fact, any reaction that involves the capture of protons by light elements will be too fast, and any reaction that involves the capture of protons by heavy elements will be too slow.

The two key reactions we now take to comprise stellar nucleosynthesis in main sequence stars, the pp chain \citep{BetheCritchfield1938} and the CNO cycle \citep{Bethe1939,Weizsacker1939}, are much more promising candidates for stellar energy production precisely because they have the right sort of reaction rates to match inferred stellar lifetimes and produce no extra isotopes as by-products, other than the hydrogen-to-helium transition, to match observed isotope abundances. Moreover, terrestrial measurements of nuclear reaction cross-sections indicate that the CNO cycle is highly temperature-sensitive, much more so than the pp chain, and for peak efficiency requires temperatures higher than the core of stars such as our Sun. Thus, for such stars, the pp chain is the main contributor to energy production, and the CNO cycle gains precedence in much larger stars.

\subsection{Inductive Triangulation}
Let us consider the structure of the relevant inferences using the philosophical toolkit we have developed above. As already noted, photonic phenomena relating to stellar nucleosynthesis are unmanipulable and inaccessible. The observational and experimental evidence that we have thus described is only able to support the theory of the phenomena based upon quite complicated modes of inference. For the first constraint, measurements of isotope abundances consist of the observation of spectra from both stellar surfaces and in interstellar space, which are cross-referenced to terrestrially observed spectra. We thus have two types of source phenomena: first, phenomena of the stellar surfaces and phenomena in interstellar space, each of which are accessible but not manipulable (\textbf{b}); and, second, the terrestrial atomic systems that are experimented upon to measure their spectra (\textbf{a}). The inference from these source systems to our target system, nuclear reactions in the stellar interior, is then validated via a range of independently established theories. In particular, theories relating to the origin of interstellar matter in both the explosion of stars via supernovae and from the big bang and, moreover, the atomic structure of elements; although it is worth noting that the two complementary theories of the origin of interstellar matter are themselves partly justified by an empirically adequate model of nucleosynthesis. It is of course hugely significant here that in such inferences we must assume that the experiments to determine the spectra of terrestrial isotopes are probative of stellar surface and interstellar isotopes. This is precisely the intra-type uniformity assumption that we have discussed extensively already.

For the second constraint, stellar core temperatures and lifetimes are attained from the inferred relationship between stellar mass and surface temperature owing to the standard interpretation of the Hertzsprung-Russell (HR) diagram. The HR diagram is a plot of observed luminosity against effective temperature and provides a model of stellar evolution, which itself, as above, relies on assumptions about stellar nucleosynthesis. Given the narrative of stellar evolution derived from the HR diagram, observations of relative stellar luminosities in globular clusters, which contain stars assumed to be all of the same age, can provide good estimates for the sorts of time scales that stars of different masses live. These astronomical observations can then be complemented with laboratory evidence (usually from particle accelerators) for nuclear reaction rates and cross-sections to provide constraints on stellar core temperatures and stellar lifetimes. These in turn place constraints on proposed nuclear reactions in the stellar core and thus the empirical viability of models of stellar nucleosynthesis. Once more we have two types of `source' phenomena: astrophysical observations of phenomena that are accessible but not manipulable (\textbf{b}); and the terrestrial nuclear phenomena that are manipulated in particle accelerators (\textbf{a}). And once more inferences from these source systems to our target system, the interior of stars, is then validated via a range of independently established theories. It is important to emphasise that at the heart of this chain of reasoning is the intra-type uniformity assumption, as before, but also the spatial (and, by extension, temporal) uniformity of the strong and weak nuclear force determining the nuclear reaction rates. In addition, there is a further appeal to temporal uniformity in assuming that the observed stars are tokens of the same types as the stars in the past that were the progenitors of the interstellar matter. It is precisely this use of multiple independently justifiable lines of inductive support, mutually supporting the overall inductive inference to the form of the nuclear reactions in the stellar interior, that we call inductive triangulation. Without these uniformity principles, the inference from the terrestrial to astrophysical phenomena could not be justified. Ultimately, if these inferences are doubted, both of these lines of evidence can be called into question. Thus, the non-rule-circular defence of the intra-type uniformity assumption is an essential `backstop' against the reasonable form of inductive scepticism we have discussed earlier.

Together we take these two lines of evidence to be sufficient to categorise the model of stellar nucleosynthesis as well supported. That is, given such evidence, scientists are justified in treating the empirical consequences of the model as likely to be true, in the relevant domain. There are, however, plausibly still reasons to doubt the model, in part because of the level of background theory mediation and the lack of empirical access to photonic phenomena in the interior of stars.

The final piece of evidence, that we take to establish the model as empirically viable in its domain beyond reasonable doubt, involves a means of gaining access to non-photonic phenomena in the interior of stars. However, once more, crucially this evidence is only in fact able to support the model when combined with terrestrial experiments, and so is naturally underpinned by uniformity principles. We have made a point so far of the fact that obtaining direct photonic observational evidence of the nuclear reactions in stellar cores is not possible. However, on account of the fact that the neutrinos produced in the nuclear reactions in the stellar core interact so weakly with matter, it is highly probable for them to escape the star without interacting, allowing us to detect on Earth stellar neutrinos directly from the stellar core. The neutrino flux from the reactions in the interior of the Sun can be observed at Earth and compared to the theoretical value of neutrino flux deduced from the theorised energy production process in the solar core. Fascinatingly, the quantitative correspondence desired did not obtain when the first solar neutrino detection experiments where conducted \citep{Davis1968}. Rather, it is only after the hypothesis of neutrino oscillations that solar neutrino experiments sensitive to the different neutrino flavours could be devised \citep{Ahmad2001}. With these solar neutrino experiments, along with subsequent terrestrial neutrino experiments, the predicted solar neutrino flux could be corrected and the correspondence between observation and prediction obtained \citep{Liccardo2018}. Our story here is thus partially modified from the above. In this case, the target phenomena are neutrino reactions in the stellar core which are accessible but not manipulable (\textbf{b}). The source phenomena are solar and atmospheric neutrinos, as well as neutrinos in terrestrial accelerators, that are manipulable (\textbf{a}) and have been established as displaying oscillation. Again, there is an appeal to intra-type uniformity -- neutrinos on Earth are like neutrinos in the stellar core in the relevant respects -- and spatial uniformity -- the weak nuclear force is invariant under spatial translations -- at the heart of the reasoning. This provides a further independently justifiable line of inductive support for the inference to the form of the nuclear reactions in the stellar interior, in accordance with inductive triangulation.

In summary, the model of stellar nucleosynthesis provides an example of target phenomena that are unmanipulable and inaccessible (\textbf{c}), the theory for which was well supported before means of access via neutrino experiments were found. Plausibly, it is this latter access that established the model as empirically viable in its domain, beyond reasonable doubt. However, this evidence, like the earlier evidence, relies crucially upon inferences from the terrestrial to the astrophysical grounded upon inductive evidence for intra-type uniformity, supported by inductive triangulation. Moreover, the realm of phenomena established for stellar nucleosynthesis includes inaccessible target phenomena such as that relating to photons in the interior of stars. This point is of particular significance in the context of black holes and analogue experiments considered in the next section.  

\section{Case Study II: Hawking Radiation in Analogue Black Holes} 
\label{sec:Hawking}
\subsection{The Argument for Confirmation}
Hawking radiation \citep{hawking:1975} is a thermal phenomenon that is predicted to be generically associated with black holes. In practice, it is impossible to obtain direct experimental evidence of Hawking radiation in astrophysical black holes. This is because for astrophysical black holes the temperature is vastly smaller than the cosmic microwave background, and so most likely outside the range of even the most fantastically sensitive future telescopes. Despite the absence of any direct experimental evidence, Hawking radiation is widely believed to be actual by theoretical physicists on the basis that it is supported by various lines of theoretical argument.\footnote{A extensive summary of the various theoretical arguments is provided in \citep{wallace:2018}. See \citep{Gryb:2018} for an overview of outstanding theoretical reasons to worry about the status of Hawking radiation.}

Not long after the original derivation of Hawking radiation, it was proposed by Unruh that a similar thermal effect might exist in the context of sound in fluid systems \citep{unruh:1981}. In particular, Unruh showed that the key elements of Hawking's calculation could be re-applied in the context of a semi-classical model of sound in fluids. An alternative medium for constructing acoustic black holes, that obeys equations of the same form as those of a fluid in an appropriate limit, is given by a Bose-Einstein Condensate (BEC) \citep{garay:2000}. There are now a huge number of potential analogue realisations of the Hawking effect: phonons in superfluid helium-3, `slow light' in moving media, travelling refractive index interfaces in nonlinear optical media, laser pulses in nonlinear dielectric media \citep{jacobson:1998,philbin:2008,belgiorno:2010,unruh:2012,liberati:2012,nguyen:2015,Jacquet:2018}. Recent years have seen a proliferation of experiments designed to probe the phenomenon of Hawking radiation via analogue black hole systems. Reports on these experiments include claims of the observation of classical, thermal aspects of Hawking radiation in an analogue white hole created using surface water waves \citep{weinfurtner:2011,weinfurtner:2013} and experiments leading to the observation of the quantum effect via the correlation spectrum of entanglement across an acoustic horizon in a BEC \citep{steinhauer:2016,deNova:2019}.\footnote{For more on surface water wave experiments see \citep{rousseaux:2008,rousseaux:2010,michel:2014,unruh:2014,euve:2016,torres:2017,euve:2020}. For further results and discussion of Steinhauer's BEC experiments see \citep{steinhauer:2014,steinhauer:2015,Finke:2016,steinhauer:2016a,Nova:2018,Leonhardt:2018}.}

In general, in analogue experiments we designate the `source' phenomena as that which is displayed by the type of physical system that is manipulated by the experimenter. The `target' phenomena is then that which is displayed by the type of physical system to which the `source' system stands in analogy.\footnote{This is the standard and well-established philosophical terminology for discussions of analogical reasoning in science \citep{hesse:1963,bailer:2009,Bartha:2013}.} In the case of an analogue experiment on Hawking radiation, this means that the target phenomena is Hawking radiation in astrophysical black holes. This is clearly an unobservable phenomenon of the unmanipulable, inaccessible type (\textbf{c}). It has been claimed in the literature that analogue experiments can in principle provide inductive support for the theoretical models of such phenomena on the basis of external validation via `universality arguments' \citep{Dardashti:2015,Dardashti:2019,Thebault:2016}.\footnote{This account of `confirmation via analogue simulation' draws heavily from the literature on the philosophy of computer simulation, in particular the work of  \cite{winsberg:1999,winsberg:2009,winsberg:2010}. Subsequent analysis has included extensions in terms of formal frameworks for confirmation theory \citep{Dardashti:2019,Gebharter:2019}, further exploration of the connection to conventional experiments and computer simulations \citep{Boge2018} and a contestable claim of there being a problematic circularity in the argument \citep{crowther:2019}. We will return to this last point of controversy shortly. An excellent overview which includes discussion of many relevant issues can be found in \cite[\S5.1]{Bartha:2013}.} The paradigmatic model of such arguments is the analysis of Unruh and Sch\"utzhold  \citep{unruh:2005} who provide theoretical reasons to expect that, under certain conditions, any modifications to the Hawking flux by high energy modes will be negligible.\footnote{For further work on these issues, using a range of different methodologies, see for example \citep{corley:1998,himemoto:2000,barcelo:2009,coutant:2012}. For discussion of the parallel trans-Planckian problem in analogue cosmology see \citep{cha:2017}.} Unruh and Sch\"{u}tzhold show that a wide family of trans-Planckian effects can be factored into the calculation of Hawking radiation via a non-trivial dispersion relation. To lowest order and given certain modelling assumptions, Hawking radiation, both astrophysical and acoustic, is independent of the details of the underlying physics.  A significant distinction that can be made in this context is between robustness and universality \citep{batterman:2000,Gryb:2018}. Robustness is the insensitivity of a phenomenon under a token-level variation with respect to different possible micro-physics in a single type of system. Universality is the insensitivity of a phenomenon under a type-level variation between systems with fundamentally different material constitution (e.g. BECs and a classical fluid). Given these definitions, we can plausibly take the work of Unruh and Sch\"{u}tzhold to be an argument for both the robustness and the universality of the Hawking effect.

The argument for inductive support for the model of black hole Hawking radiation based upon analogue experiments validated via universality arguments thus has a very similar form to that for other inferences about inaccessible astrophysical phenomena, such as the case study in \S\ref{sec:stellar}. In particular, we have a reliance on a principle of uniformity between a manipulable unobservable phenomenon in a source system (analogue Hawking radiation) and an inaccessible unobservable phenomenon in a target system (black hole Hawking radiation). Such an inference closely parallels that between, for instance, nuclear processes in terrestrial particle accelerators and in the interior of stars. In particular, with regard to photonic processes at least, the interior of a star is inaccessible for precisely the same reason as the event horizon of a black hole: in both cases the relevant flux of photons is vanishingly small. The contrast is that whereas in more conventional experiments the source system is of the same type as the target system, here the reliance is on the source being in the same universality class as the target system. Thus, there is, \emph{prima facie}, only a fundamental difference between the two forms of inference if one thinks that there is a fundamental difference in kind between intra-type regularity principles, as embodied by natural kind arguments, and inter-type regularities, as embodied by universality arguments.

\subsection{The Circularity Claim}

It will now prove worthwhile to consider a recent attempt to undermine this argument for the inductive support of models of black hole Hawking radiation via the combination of universality arguments and analogue experiments. In particular, \cite{crowther:2019} have claimed that to make such an argument one must assume ``the physical adequacy of the modelling framework used to describe the inaccessible target system'' and that this implies that ``arguments to the conclusion that analogue experiments can yield confirmation for phenomena in [inaccessible] target systems, such as Hawking radiation in black holes, beg the question'' (p.~1). As stated, it is a little difficult to know what to make of the argument of Crowther \emph{et al}. In particular, they talk about ``begging the question'' as the ``inductive analogue'' of the deductive fallacy of the same name. At some points it appears that they mean to indicate the premise circular version of circular reasoning since they explicitly talk about the problem being that one ``assume[s] the conclusion that [one] is trying to establish'' (p.~20). However, as we have seen already, as an ampliative inference, inductive reasoning simply cannot be premise circular: all premise circular arguments are non-ampliative by definition. Thus, it would be simply incoherent to claim that an inductive argument for confirmation via analogue simulation is premise circular when `premise circular' is given its standard interpretation. An argument simply cannot be both inductive and premise circular when each term is standardly understood.

Is there a coherent sense in which we can construct an argument as both inductive and `premise circular' in some analogous but different sense to the standard definition that `the conclusion is listed amongst the premises'?  Let us consider what Crowther \emph{et al.} say explicitly to see if we can discern what they might mean by the `inductive analogue' of premise circularity. Consider the following statement in the context of the informal arguments for confirmation via analogue simulation in the black hole case provided by  \cite{Dardashti:2015} and \cite{Thebault:2016}:
\begin{quote}  
	\ldots by assuming that black holes are accurately described by the modelling framework from which the derivation of Hawking radiation is a necessary consequence, [Dardashti \emph{et al.} and Th\'{e}bault] already assume the conclusion they are trying to establish---that Hawking radiation exists in black holes. It is in this sense that they are begging the question. (p.~20)
\end{quote}
Then, in the context of their analysis of the Bayesian argument towards confirmation via analogue simulation due to \cite{Dardashti:2019} they say:
\begin{quote}
	\ldots the circular dependence of conclusion on premise remains, as it must still be presupposed that black holes are the kind of system that, with at least some non-zero probability, exhibit certain physical behaviour, which is precisely what one seeks to establish with analogue confirmation. (p.~22)
\end{quote}
We might thus plausibly interpret Crowther \emph{et al.} to take that the `premise circularity' in question is that it must be assumed in the informal arguments that the semi-classical black hole modelling framework, including Hawking radiation, is `adequate' as a description of actual black holes or that, in the formal arguments, that there is a non-zero probability that black holes exhibit Hawking radiation.
 
Given this interpretation, \textit{prima facie}, it might appear that Crowther \emph{et al.} have identified a form of structural problem in the arguments in question that is an inductive analogue of premise circularity. However, a little consideration of the nature of inductive reasoning in general, and Bayesian reasoning in particular, shows that this putative line of criticism either reduces to inductive scepticism or deductive premise circularity. Inductive arguments are arguments towards a relation of non-deductive inferential support between evidence and a hypothesis. They have the general feature that the hypothesis must be assumed to be possible for them to function -- after all, this is what we standardly mean by `hypothesis'. Informally, if one is certain that the hypothesis ``all ravens are black'' is false, since for instance one has seen a white raven, then no matter how many black ravens one then sees, one cannot take the hypothesis to be inductively supported. Formally, if we assign a hypothesis regarding the empirical adequacy of a theory a prior probability of zero, then no evidence can ever lead to confirmation in Bayesian terms since the posterior probability will always remain zero under Bayesian conditioning.\footnote{Explicitly, following \citep{DardashtiHartmann:2019}, consider the propositions, $T$: ``the theory under-consideration is empirically adequate''; and $E$: ``empirical evidence in favour of the theory obtains''. Consider an agent with a prior probability distribution $P$ over the propositional variables \textit{T} and \textit{E} (with values $T$ and $\neg T$ and  $E$ and $\neg E$ respectively). Bayesian updating implies that the posterior probability distribution $P^\star$ for the probability of $T$ is given by:
\begin{equation}
	P^\star(T)=\frac{P(E|T)P(T)}{P(E)}.
\end{equation}
This means that if $P(T)=0$ then we necessarily have that $P^\star(T)=0$ (or is ill-defined) and thus confirmation is impossible.}

Assuming some minimal adequacy of the modelling framework as a description of a target system, or a non-zero probability for the relevant hypothesis, is a \textit{necessary pre-condition of all inductive reasoning}. If \textit{that} feature is what we should understand as the inductive analogue of `premise circularity', then Crowther \emph{et al.}'s argument immediately descends into a sceptical argument against inductive knowledge in general based upon unreasonable doubt. Conversely, if the word `precisely' is taken literally in the second quote above, and the issue is understood to be that the conclusion of the argument is identical to the proposition that `there is a non-zero probability that black holes exhibit Hawking radiation', then there would indeed be premise circularity but only in the standard deductive sense of the term. And such a reading is clearly implausible given that the entire discussion is framed by ideas of confirmation and inductive evidence. We can thus conclude that there is no viable way to parse the relevant notion of an `inductive analogue' of premise circularity in coherent terms that avoids collapse into inductive scepticism and unreasonable doubt.

A more promising reconstruction of Crowther \emph{et al.}'s argument, and one with which we expect them to agree, is available via reference to the idea of rule circularity. The reconstruction would run as follows. In order to externally validate the inference from source to target system via universality arguments, one must make inferences based upon a uniformity principle of a novel kind: that is, inter-type uniformity between accessible and inaccessible phenomena. However, it is the reliability of precisely such a uniformity principle that is itself in question when we are trying to ascertain external validity of analogue experiments. Hence there is indeed a form of rule circularity implicit in the argument for confirmation via analogue simulation. We take this to be the feature of the external validation of analogue experiments that Crowther \emph{et al.} find worrying.

Recall, however, that whether or not rule circularity is dialectically undermining depends upon the context of the debate. If one is arguing with an inductive sceptic, then appeal to inductive triangulation may enable one to convert the putative problem with rule circularity into one of justificatory regress. However, it is not clear that this improves the anti-sceptic's position in the argument: it is not clear that they have provided any dialectically convincing response to the inductive sceptic. On the other hand, since Crowther \emph{et al.} are not assuming the position of inductive sceptics, and so are not offering an argument based upon unreasonable doubt, they surely cannot rule out strategies for inductive triangulation that lead to a termination of their justificatory demands. In particular, their explicit argument suggests that they evidently \textit{do} admit as unproblematic at least some inductive inference, and thus the path is open for dialectically convincing mitigation of their reasonable doubts via inductive triangulation.

\subsection{Inductive Triangulation}

Consistent between the various accounts of confirmation via analogue simulation is the claim that we require a combination of multiple independent analogue experiments to be successfully performed to support the case for black hole Hawking radiation \citep{Dardashti:2015,Dardashti:2019,Thebault:2016}. This is equivalent to an enumerative mode of inductive reasoning between accessible systems of different types. In performing multiple successful analogue experiments one is providing inductive evidence for inter-type uniformity between different accessible phenomena. Consider a hypothetical dialectic with a selective sceptic regarding inductive reasoning based upon inter-type uniformity. Assume such a sceptic admits the application of enumerative induction. Then one can provide dialectically convincing evidence by demonstrating instances of successful inter-type uniformity: that is, experimentally demonstrating some phenomena in one type of system, inferring that the phenomena exist in a second type of system, and then demonstrating the phenomena in the second system. There is nothing special about inter-type inductive reasoning that bars inductive triangulation of the usual form. Furthermore, as Crowther \emph{et al.} themselves note, what counts as the same `type' or `kind' of system is to some extent context dependent. Thus, a lot of intra-type reasoning in science might be reinterpreted as inter-type reasoning in a given context (we would expect this to be particularly true in the life sciences, and also in molecular and materials science). Moreover, there exist examples of scientists using Wilsonian type universality arguments to justify inter-type reasoning of this form in a range of condensed matter contexts \citep{thouless1989,prufer:2018,erne:2018,eigen:2018}. Thus inter-type inductive reasoning is in fact far more credible, and mundane, than it might at first seem.

Let us then instead interpret the scepticism to be regarding the rule of inference from accessible to inaccessible phenomena. That the `inaccessibility' of the target system is at the heart of Crowther \emph{et al.}'s worry is indicated numerous times in the paper. In particular, inaccessibility of the target system is referred to as both the ``key difference'' between analogue and conventional experiments (p.~7) and, moreover, the reason why analogue experiments \textit{in general} are taken by Crowther \emph{et al.} to not be potentially confirmatory (p.~24). But here it is worth keeping in mind our analysis of stellar nucleosynthesis above. Clearly there are inaccessible target systems about which we can formulate theories and models that can be well supported by combinations of different lines of evidence. Inaccessibility of the target system is thus \textit{not} the key difference between analogue and conventional experiments since conventional experiments can and do licence claims regarding inaccessible target phenomena. To exclude confirmation of inaccessible target systems in principle would be to eliminate a variety of well-supported and established theories and models in contemporary physics. Furthermore, clearly if such a scepticism is voiced, it can be mitigated in a dialectically convincing way via inductive triangulation based on a vast array of successful scientific practices. Reliable inferences concerning conclusions about inaccessible target systems are commonplace in science and thus there are abundant resources available to mitigate the concerns of a (reasonable) sceptic regarding accessible to inaccessible inductive inferences in general. Finally, it is worth noting that black holes are accessible when seen as classical systems since  empirical access to classical black hole phenomenology can be gained via the relevant gravitational interactions. For example, between the 1970s and early 2000s, astrophysical observations lead to the near unequivocal demonstration that the centre of the Milky Way holds a supermassive black hole based on the motions of nearby orbiting stars \citep{genzel:1997,ghez:2000,Curiel:2019}.\footnote{For further discussion of the fascinating methodological and epistemological problems associated with observation of black holes in astronomy see \citep{collmar-et-panel-proof-exist-bhs,hehl-et-bh-theory-obs,eckart-et-superm-bh-good-case}. Thanks to Erik Curiel for advice on this literature.} Furthermore, many of the key classical features of black holes, such as event horizons, are implied by general relativity, which is extraordinarily well confirmed. So, in fact, the inference that we are considering is from an accessible and manipulable system to an unmanipulable and \textit{partially} inaccessible system. Structurally this is a similar form of inference to that found in our stellar nucleosynthesis case.

This latter point is telling for the more speculative claim for which we argue in this work; namely, that we take inductive triangulation to allow for the possibility of analogue experiments to play a role, when combined with appropriate conventional experimental results, in \emph{establishing} new theories. In the stellar nucleosynthesis case above, we saw explicitly how a theoretical framework characterising a set of unmanipulable and partially inaccessible phenomena could come to be regarded as established by the scientific community on account of a series of independent lines of evidence employing independent inductive inferences (that is, by inductive triangulation). We characterised the model of stellar nucleosynthesis as merely well supported as a result of the `photonic' evidence, from observed and calculated isotope abundances and observed and inferred stellar luminosities and temperatures. However, when combined with independent evidence from neutrino experiments we take the model of stellar nucleosynthesis to be established. Granted, analogue experiments showing Hawking radiation cannot by themselves establish astrophysical Hawking radiation. But the inter-type inductive reasoning from source to target employed by such experiments stands as a line of inductive inference that has the potential to be combined with other independent lines made available by future experiments, whether analogue or conventional, through the process of inductive triangulation, to conceivably establish the relevant astrophysical theories. We take the difference between the stellar nucleosynthesis case and the analogue black hole case to be a difference in degree rather than a difference in kind. 

One final option for interpreting the `circularity' claim is that Crowther \emph{et al.} might take there to be an inferential problem specific to inferences based upon the \textit{combination} of inter-type and accessible-inaccessible uniformity. The idea might be that there is some problem inherent to this combined inference that is not reducible to either of its components. It is difficult, however, to see how the logic of such reasoning would work. If it is permissible to reason separately from accessible to inaccessible systems and from systems of one type to systems of another, then we need some reason why it is not permissible, in principle, to reason from accessible systems of one type to inaccessible systems of another type. So long as our inferences obey classical logic such concatenation must be accepted. Moreover, such a form of sceptical argument would again seem liable to descent into inductive scepticism. If concatenation of distinct forms of justified inductive reasoning were not itself automatically justified, as is assumed in inductive triangulation, then it would not be difficult to generate a justificatory regress, given the possibility to `fine grain' principles of uniformity to an essentially infinite degree. Thus, again, there seems no room for reasonable doubt.

The proceeding paragraphs have provided an exhaustive analysis of all plausible grounds for a structural inferential problem that might be termed `circularity'. However, abandoning the circularity charge, a determined sceptic might wish to put forward a specific scepticism regarding the universality arguments for Hawking radiation in particular. Might these arguments not be uniquely problematic: either since they are not suitably independent from the semi-classical models which they are deployed to partially justify or have some other fundamental physical or mathematical problems that make them implausible.\footnote{Thanks to Grace Field for extensive discussions on each these points.} First, it is true that many of the same ingredients (e.g. calculating flux via Bogoliubov coefficients, or the assumption of no back-reaction of quantum scalar field on classical spacetime) that go into the Unurh-Sch\"{u}tzhold universality arguments also went into the original Hawking derivation. However, the two formal arguments also have significant differences. In particular, the universality argument is specifically about the breakdown of the semi-classical regime within which the Hawking calculation works (this is precisely what the modifications of the dispersion relation are designed to model). Moreover, there are various families of other derivations of Hawking radiation, almost completely detached from the specific details of the universality arguments. It is therefore difficult to argue that the two arguments are problematically interdependent. Second, it is certainly the case that there are reasonable grounds for scepticism regarding the universality arguments as they currently stand \citep{Gryb:2018}. However, it is important to note that the relation of inductive support between analogue experiments and astrophysical Hawking radiation only relies upon the universality arguments having non-trivial (i.e. probability neither zero nor one) credence \citep{Dardashti:2019}. Whilst one could certainly justify setting a relatively low credence, it is surely just as unreasonable to believe them to be certainly false as it is to believe them to be certainly true. There are a number of independently well-established general features that analogue black holes and astrophysical black holes share, in particular event horizons and a continuum limit. Thus some \emph{prima facie} plausibility must surely be granted to the universality arguments. Assigning a low credence in the universality arguments would mean that inductive evidence from analogue experiments cannot render conclusions about astrophysical black holes `well supported', no matter how many such experiments are carried out. However, such scepticism does not block the relation of inductive support \textit{per se}. It is thus difficult to resist the conclusion that, \textit{pace} Crowther \emph{et al.}, analogue experiments can in principle provide support for (and thus stand in confirmation relations to) theories and models describing inaccessible target systems, like black holes. 

In summary, to rule out inductive support for astrophysical Hawking radiation based upon analogue experiments is unreasonable. However, for this support to be strengthened, and reasonable doubts mitigated, both stronger universality arguments and a new generation of analogue experiments showing Hawking radiation in diverse media are needed.  

\section{Conclusion} 

The foregoing arguments and analysis notwithstanding, even if a wide range of analogue experiments were successfully conducted and the relevant universality arguments significantly strengthened, black hole Hawking radiation would certainly not be something that is beyond reasonable doubt (or `conclusively confirmed').\footnote{This is entirely consistent with the accounts provided in the literature \citep{Dardashti:2015,Dardashti:2019,Thebault:2016}. It should be noted however, that a pre-print version of \citep{Thebault:2016} (quoted by Crowther \emph{et al.}) contains an unfortunate typographical error that has introduced confusion on precisely this point.}  The probative value of the next generation of analogue experiments in part depends upon scientists' ability to combine them with other analogue experiments and universality arguments to develop a stronger case of inductive triangulation. In this context, it is worth noting that there are formal arguments \citep{Dardashti:2019} which imply that the more certain we are about the adequacy of the analogue model that describes the source system that we are experimenting on, the less effective is the evidence obtained there in confirming the adequacy of the model of the target system. We learn more about the target system by conducting future analogue experiments using media about which we are \textit{less certain} regarding their fundamental physics. Thus, by strengthening the universality arguments, \textit{and} testing them more stringently across analogue platforms within which they may break down, we can increase the inductive support for analogue Hawking radiation such that the relevant theory might plausibly be taken to be well supported. Looking further to the future, inductive triangulation allows for the possibility of analogue experiments to play a role, \textit{when combined with appropriate conventional experimental results}, in establishing new theories.

\vspace{5mm}
\singlespacing
{\footnotesize \noindent \textbf{Acknowledgments:} We are greatly appreciative to Simon Allz\'{e}n, Ana-Maria Cre\cb{t}u, Erik Curiel, Tamara Davis, Richard Dawid, Grace Field, Sebastian Lutz, Michela Massimi, Luca Moretti, and Samir Okasha for extremely helpful discussions and written comments on a draft manuscript, and to audiences in Wollongong, Geneva, London, and Bristol for valuable feedback. We are also appreciative of comments from two anonymous referees that we found very helpful in refining our arguments. Finally we are both very grateful to the hospitality of the EIDYN Centre at the University of Edinburgh, and in particular to Michela Massimi for hosting us during work on this paper.

PWE's work on this paper was supported by the University of Queensland and the Australian Government through the Australian Research Council (DE170100808). KT's work on this paper was supported by the Arts and Humanities Research Council, UK (AH/P004415/1).

}

\bibliographystyle{timebib}
\bibliography{ConfBib,dumb}

\end{document}